\documentclass[apj]{emulateapj}

\def\hMsun{h^{-1} \; M_\odot}
\def\hMpc{h^{-1} \; {\rm Mpc}}
\def\hkpc{h^{-1} \; {\rm kpc}}
\def\VhMpc{h^{3} \; {\rm Mpc}^{-3}}

\def\ergs{{\rm erg \; s^{-1}}}

\newcommand{\appropto}{\mathrel{\vcenter{
  \offinterlineskip\halign{\hfil$##$\cr
    \propto\cr\noalign{\kern2pt}\sim\cr\noalign{\kern-2pt}}}}}

\shorttitle{The HOD of X-Ray AGN}
\shortauthors{Richardson et al.}

\begin{document}

\title{
The Halo Occupation Distribution of X-ray-Bright Active Galactic Nuclei: A Comparison with Luminous Quasars
}

\author{Jonathan Richardson$^{1}$, Suchetana Chatterjee$^{2,3}$, Zheng Zheng$^{4}$, Adam D.\ Myers$^{2}$, and Ryan Hickox$^{5}$}
\affiliation{$^{1}${Department of Astronomy and Astrophysics, University of Chicago,
			Chicago, IL 60605, USA}\\
             $^{2}${Department of Physics and Astronomy, University of Wyoming,
			Laramie, WY 82072, USA}\\
             $^{3}${Yale Center for Astronomy and Astrophysics, Department of Physics, Yale University, 
			New Haven, CT 06520, USA}\\
             $^{4}${Department of Physics and Astronomy, University of Utah,
			Salt Lake City, UT 84112, USA}\\
             $^{5}${Department of Physics and Astronomy, Dartmouth College,
			Hanover, NH 03755, USA}\\
	    }
\email{jonathan.richardson@uchicago.edu}
\email{schatte1@uwyo.edu}

\begin{abstract}

We perform halo occupation distribution (HOD) modeling of the projected two-point correlation function (2PCF) of high-redshift ($z \sim 1.2$) X-ray-bright active galactic nuclei (AGN) in the XMM-COSMOS field measured by \citeauthor{allevatoetal11} The HOD parameterization is based on low-luminosity AGN in cosmological simulations. At the median redshift of $z\sim1.2$, we derive a median mass of $1.02_{-0.23}^{+0.21}\times 10^{13} \; \hMsun$ for halos hosting central AGN and an upper limit of $\sim10\%$ on the AGN satellite fraction. Our modeling results indicate (at the 2.5$\sigma$ level) that X-ray AGN reside in more massive halos compared to more bolometrically luminous, optically-selected quasars at similar redshift. The modeling also yields constraints on the duty cycle of the X-ray AGN, and we find that at $z\sim1.2$ the average duration of the X-ray AGN phase is two orders of magnitude longer than that of the quasar phase. Our inferred mean occupation function of X-ray AGN is similar to recent empirical measurements with a group catalog and suggests that AGN halo occupancy increases with increasing halo mass. We project the {\it XMM}-COSMOS 2PCF measurements to forecast the required survey parameters needed in future AGN clustering studies to enable higher precision HOD constraints and determinations of key physical parameters like the satellite fraction and duty cycle. We find that $N^{2}/A \sim 5\times 10^{6}$ deg$^{-2}$ (with $N$ the number of AGN in a survey area of $A$ deg$^2$) is sufficient to constrain the HOD parameters at the 10\% level, which is easily achievable by upcoming and proposed X-ray surveys.

\end{abstract}

\keywords{dark matter, galaxies: nuclei, large-scale structure of the universe, AGN: general}

\section{Introduction}

Galaxies are believed to reside in dark matter halos and as such make excellent probes of structure formation in the universe \citep[e.g.,][]{w&f91, kauffmannetal93, nfw95, m&w96, kauffmannetal99, springeletal05a}. Observations suggest that there exists a supermassive black hole at the center of every massive galaxy \citep[e.g.,][]{soltan82}, and hence a physical connection between black holes and their host dark matter halos is naturally expected. The black hole-halo relationship has been widely studied in both semi-analytic models and cosmological hydrodynamic simulations \citep[e.g.,][]{k&h00, w&l03, marconietal04, cattaneoetal06, crotonetal06, hopkinsetal06, lapietal06, shankaretal04, dimatteoetal08, b&s09,  volonterietal11, c&w13}. The spatial clustering of black holes can be used to probe this relationship and to obtain constraints on the formation and evolution of supermassive black holes and their role in galaxy formation. 

Clustering analyses of different types of active galactic nuclei (AGN) have been widely studied in the literature \citep[e.g.,][]{croometal04, porcianietal04, croometal05, gillietal05, myersetal06, myersetal07a, myersetal07b, coiletal07, shenetal07, wakeetal08, shenetal09, rossetal09, coiletal09, hickoxetal09, hickoxetal11, allevatoetal11, donosoetal10, krumpeetal10, cappellutietal12, whiteetal12, shenetal12a, krumpeetal12, mountrichasetal13, koutoulidisetal13}. Many of these studies employ samples of optically-bright quasars, which provide large sample sizes and accurate redshifts that enable  precise measurements of the relationship between the growth of black holes and their parent dark matter halos over a wide range in redshift. However, while quasar clustering has provided a wealth of information on the cosmological evolution of black holes, samples of optical quasars preferentially probe massive black holes accreting at high Eddington rates \citep{hopkinsetal09a, kellyetal10}. To fully understand the growth of black holes over cosmic history, we require a systematic study of the relationship between black holes and their host halos spanning a larger range of redshifts, luminosities, and black hole masses. This can be achieved through X-ray surveys.

AGN clustering is most often studied using the two-point correlation function \citep[2PCF; e.g.,][]{arp70}. For a given cosmological model, the amplitude of the large-scale AGN 2PCF relative to the dark matter 2PCF (i.e., the square of the AGN bias) can be estimated. From the value of this bias factor an estimate of the typical mass of AGN-hosting dark matter halos can then be obtained \citep[e.g.,][]{jing98,shethetal01}. Previous studies suggest that, compared to optical quasars, X-ray-selected AGN are more strongly clustered and reside in more masive host halos \citep[e.g.,][]{gillietal05, coiletal09, hickoxetal09, allevatoetal11, mountrichasetal13,cappellutietal12}. However, the host halo mass range is insufficiently constrained for a definitive conclusion \citep[for a recent review, see][]{cappellutietal12}.

Beyond simple bias measurements, a powerful analytic technique known as the halo occupation distribution (HOD) formalism  \citep[e.g.,][] {m&f00, seljak00, b&w02, zhengetal05, z&w07} can be used to investigate the full relationship between AGN and their host halos. The HOD is characterized by the probability $P(N|M)$ that a halo of mass $M$ contains $N$ AGN of a given type, together with the spatial and velocity distributions of AGN inside individual halos. As long as the HOD, at a fixed halo mass, is statistically independent of the large scale environments of halos \citep[e.g.,][]{bondetal91}, it provides a complete description of the relation between AGN and halos, allowing the calculation of any clustering statistic (e.g., N-point correlation function, void probability distribution, pairwise velocity distribution) at all scales for a given cosmological model. This implies that if the HOD can be constrained empirically, it will encode all of the information that measured clustering properties can convey about AGN formation and evolution.

The HOD approach has been used by several authors to interpret AGN and quasar clustering measurements \citep[e.g.,][]{wakeetal08, shenetal10, miyajietal11, starikovaetal11, allevatoetal11, richardsonetal12, k&o12}. In particular \citet{richardsonetal12}, and \citet{k&o12} made clustering measurements of optically luminous quasars for both small ($<1$ Mpc) and large scales and performed HOD modeling of the 2PCF to infer the relation between quasars and their host dark matter halos at $z\sim1.4$. Similar analysis was carried out by \citet{shenetal12a} for the quasar-galaxy cross-correlation function (CCF). A natural extension of work on the HOD of optically luminous quasars is to study the black hole-halo relationship for lower luminosity AGN, selected from optical or X-ray surveys. In the current work, we use the 2PCF measurements of \citet{allevatoetal11} and a theoretically motivated model of the AGN HOD similar to that used by \citet{richardsonetal12} to interpret the spatial clustering of X-ray-selected AGN at $z \sim 1.2$. We compare our results with previous studies of the quasar HOD by \citet{richardsonetal12} and construct an evolutionary picture of AGN growth and the role of AGN in large-scale structure formation (similarly to \citealt{hickoxetal09}, but now from the HOD perspective, and at a higher redshift of $z \sim 1$).

Our paper is organized as follows. In $\S2$, we briefly describe our data sets, the parameterization of the AGN HOD, and the theoretical modeling of the 2PCF. We present the results of our HOD modeling in $\S3$. Finally, we discuss the implications of our results in $\S4$ and summarize them in $\S5$. Throughout the paper we assume a spatially flat, $\Lambda$CDM cosmology: $\Omega_{m}=0.26$, $\Omega_{\Lambda}=0.74$, $\Omega_{b}=0.0435$, $n_{s}=0.96$, $\sigma_{8}=0.78$, and $h=0.71$. We quote all distances in comoving $\hMpc$ and masses in units of $\hMsun$ unless otherwise stated.

\section{Datasets and Methodology}

In this section we briefly describe our clustering data set and methodology for the HOD analysis.

\subsection{Clustering Data Set}
\label{sec:data}

We model the projected 2PCF measurements of \citet{allevatoetal11} for AGN in the {\it XMM}-COSMOS survey. COSMOS covers $1.4 \times 1.4$ deg$^{2}$ equatorial field\footnote{centered at J$_{2000}$ (RA, Dec) = ($150.1083^{\circ}$, $2.210^{\circ}$)} and uses multiwavelength data from X-ray to radio bands. {\it XMM}-Newton surveyed $2.13$ deg$^{2}$ of the COSMOS field in the 0.5-10 kev energy band for a total of 1.55 Ms. The resulting {\it XMM} point source catalog contains $1822$ objects, $1465$ of which are {\it XMM}-COSMOS AGN detected in the soft X-ray band \citep{hasingeretal07, cappellutietal07, cappellutietal09}. 

Using the $780$ {\it XMM}-COSMOS AGN that have spectroscopic confirmations, \citet{allevatoetal11} applied a magnitude cut of $I_{{\mathrm AB}} < 23$ to obtain a clustering sample of $593$ X-ray-selected AGN. The sample spans the redshift range 0.0--4.0 with a median redshift of $1.2$. The X-ray luminosities of the AGN span the range $\sim 10^{41}$--$10^{45} \; \ergs$ with a median luminosity of $\sim 6 \times 10^{43} \; \ergs$. We refer the reader to \citet{allevatoetal11} for a detailed discussion of the 2PCF calculation.

Correcting for spectroscopic completeness, we find the redshift-averaged observational number density of the AGN to be $(6.7 \pm 0.7) \times 10^{-5} \; \VhMpc$, where the $\sim 10 \%$ uncertainty corresponds to the characteristic level of scatter we observe between the redshift averages computed by discrete summation and by fitting a decaying exponential to the observed redshift distribution (see Figure\ 1 of \citealt{allevatoetal11} for the redshift distribution). The $2.13$\,${\rm deg}^2$ survey area probes a total volume of $0.09$ Gpc$^{3}$. For this work, we compute the source density by averaging over the volume-weighted number density for each redshift bin (i.e., the number of sources within the redshift range divided by the volume of the redshift slice).

\subsection{HOD Modeling}
\label{sec:modeling}

We adopt the theoretical HOD model for AGN proposed by \citet{chatterjeeetal12}, which was inferred for lower luminosity AGN ($L_{\rm bol} \leq 10^{42} \; \ergs$) in cosmological simulations. However, \citet{richardsonetal12} have found this model to also provide an excellent interpretation of the clustering of quasars observed in the Sloan Digital Sky Survey (SDSS). We therefore apply this particular model to moderate-luminosity X-ray AGN assuming a universality in the general HOD properties of AGN. See \S\ref{sec:discussion_fit} for a detailed discussion of alternative HOD models.

This model represents the AGN mean occupation function as the sum of its physically illustrative central and satellite components, $\langle N_{\mathrm{cen}}(M)\rangle$ and $\langle N_{\mathrm{sat}}(M)\rangle$, respectively. The mean occupation function is given as the sum of a softened step function (central component) and a rolling-off power law (satellite component),  
\begin{eqnarray}
\langle N(M)\rangle & = & \frac{1}{2}\left[1+{\rm erf}\left(\frac{{\rm log} M-{\rm log} M_{\rm{min}}}{\sigma_{\rm{log M}}}\right)\right] \nonumber \\ & + & \left(\frac{M}{M_{1}}\right)^{\alpha} \exp \left(-\, \frac{M_{\rm{cut}}}{M} \right),
\end{eqnarray}
where $M_{\mathrm{min}}$, $\sigma_{\log M}$, $M_{1}$, $\alpha$ and $M_{\mathrm{cut}}$ are free parameters. In order to minimize the parameter degeneracy in our modeling, we fix $M_{\mathrm{cut}}$ to a value $\ll 10^{12} \; \hMsun$ so that the exponential factor, $\exp \left(-M_{\rm{cut}}/M \right)$, approaches unity over the range of physical halo masses. This reduces our satellite parameterization to a two-parameter power law.

Using the routine developed in \citet{zhengetal07}, we perform a Markov Chain Monte Carlo (MCMC) modeling of the projected 2PCF to sample the four-dimensional parameter space of the AGN HOD. Each MCMC chain discussed herein contains $60,000$ points in the HOD parameter space. Since the clustering sample is sparse, we calculate the $\chi^{2}$ value of each point using only the diagonal elements of the covariance matrix \citep[see, e.g., the appendixes of][]{myersetal07a,rossetal09}. Each $\chi^{2}$ value accounts for both the observed 2PCF and number density of AGN. This routine includes the effects of halo exclusion, nonlinear clustering and scale-dependent halo bias, where halos are defined as objects with a mean density of $200$ times that of the background universe. The halo mass function is computed according to the formula in \citet{jenkinsetal01} and the large-scale halo bias factor is computed using the formula in \citet{tinkeretal05}. We adopt flat priors in logarithmic space for $M_{\mathrm{min}}$ and $M_{1}$ and in linear space for $\alpha$ and $\sigma_{\log M}$ ($\sigma_{\log M} > 0$). For improved computational efficiency, we further require that $0.5 < \alpha < 4.0$. Although lower values of $\alpha$ have been reported in recent studies \citep[e.g.,][]{allevatoetal12}, we tested our modeling with a relaxed prior of $-0.5 < \alpha < 4.0$ and found our results to be insensitive to the lower limit on $\alpha$. As will be discussed in \S\ref{sec:results_current}, our modeling favors values of $\alpha \gtrsim 1$.

As in \citet{richardsonetal12}, we adopt the following conventions in our modeling. We assume that the halo occupations of central and satellite AGN are independent (i.e., the number of satellites in a given halo does not depend on whether there is a central AGN), as \citet{chatterjeeetal12} found no evidence of a correlation between the activity of central and satellite black holes in a hydrodynamic simulation. For halos of a given mass, we assume that the central occupation numbers obey a nearest integer distribution \citep{b&w02} and the satellite occupation numbers obey a Poisson distribution. This has also been found to be the case for AGN in cosmological simulations \citep{degrafetal11b, chatterjeeetal12}. We represent the spatial distribution of satellite AGN within halos as an NFW profile \citep{nfw97} with the concentration-mass relation from \citet{bullocketal01},
\begin{equation}
c(M,\,z)= \frac{c_{0}}{1+z} \left( \frac{M}{M_{*}} \right)^{\beta}, 
\end{equation}
where $M_{*}$ is the nonlinear mass for collapse at $z=0$, and $\beta=-0.13$. We adopt $c_{0}=32$, motivated by the high concentration observed for local AGN \citep[e.g.,][]{l&m07}. We have verified that our modeling only weakly depends on $c_{0}$ for a wide range of values, from $\sim10$ to $\sim60$ \citep[]{richardsonetal12}.

To enhance its statistical power, the clustering sample has been constructed over a broad redshift and luminosity range to maximize the volume and the number of sources. The 2PCF obtained in this way can be interpreted as an average over the redshift and luminosity intervals. However, extending this interpretation to the HOD could be problematic since the modeling uses halo properties (e.g., mass function, bias factor) at the median redshift. Redshift evolution of the halo properties can lead to systematic effects that exceed the statistical uncertainties reflected in the modeling results. 

Although a full interpretation of the measured 2PCF would require additional parameters accounting for evolution, \citet{richardsonetal12} have demonstrated that the HOD can be meaningfully interpreted as representing the HOD for objects at the median redshift of the sample (to within the quoted uncertainties), if the 2PCF measured over a wider redshift range around the median redshift (i.e., the average 2PCF) is statistically consistent with the true 2PCF of objects at the median redshift. Since we are not in a position to explicitly check this condition for the \citet{allevatoetal11} clustering sample, we adopt the median-redshift interpretation with the assumption that the clustering evolves weakly with redshift. See \S\ref{sec:discussion_fit} and \citet{richardsonetal12} for additional discussion regarding this chosen interpretation.

\section{Results}

We first present the constraints on the AGN HOD obtainable with current data. Then we forecast improvements in the precision of the 2PCF to estimate the requisite observational parameters to obtain high-precision (few percent-level) HOD constraints.

\subsection{Current Constraints}
\label{sec:results_current}

%%%%%%%%%%%%%%%%%%%%%%%%%%%%%%%%%%%%%%%%%%%%%%%%%
\begin{figure*}
\begin{center}
\begin{tabular}{c}
\includegraphics[width=16cm]{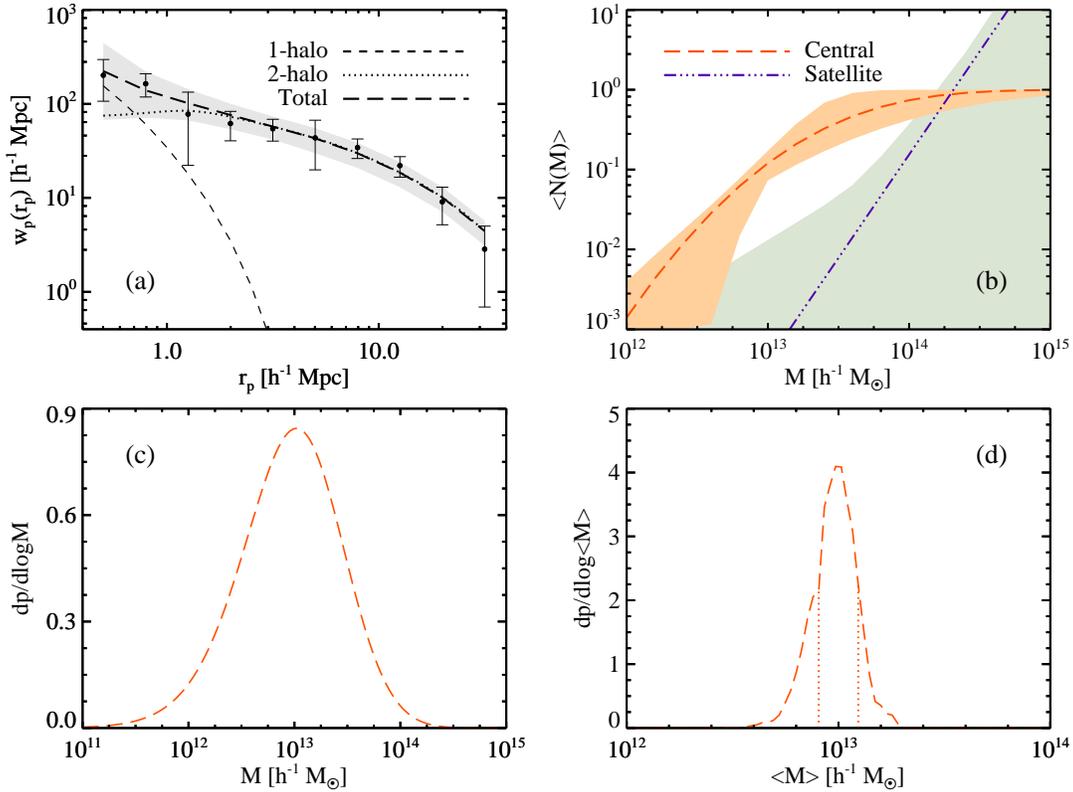}
\end{tabular}
\caption{
\label{fig:fig1}
HOD fit to the 2PCF of $z \sim 1.2$ X-ray AGN. Panel (a): the projected 2PCF of \citeauthor{allevatoetal11} (\citeyear{allevatoetal11}; data points and error bars) against the prediction of our best-fit HOD model (long dashed line) separated into its constituent 1-halo (short dashed line) and 2-halo (dotted line) terms. Panel (b): the mean occupation function of AGN, decomposed into its central (dashed line) and satellite (dot-dashed line) components. In both panels (a) and (b), the shaded envelopes indicate the $68\%$ confidence regions. Panel (c): the full probability density function of host halo masses for central AGN. This distribution is obtained by multiplying the central mean occupation function with the differential halo mass function and averaging over all the models in the MCMC chain (see \S\ref{sec:results_current} for discussion). Panel (d): the probability density function of the median mass of halos hosting central AGN. This distribution is obtained from the median host halo mass of each model in the MCMC chain. The vertical lines denote the central $68\%$ confidence interval.
}
\end{center}
\end{figure*}
%%%%%%%%%%%%%%%%%%%%%%%%%%%%%%%%%%%%%%%%%%%%%%%%%%

In Figure~\ref{fig:fig1} we show our four-parameter HOD fit for X-ray AGN at $z\sim1.2$. Panel (a) shows the projected 2PCF measurements of \citet{allevatoetal11} against the theoretical prediction of our best-fit HOD model (thick long dashed line), also shown separated into its constituent 1-halo (dotted line) and 2-halo (short dashed line) terms. We identify the best-fit model as the point in our four-dimensional parameter space associated with the global $\chi^{2}$ minimum. If we rank all of the models by their $\chi^2$ statistic in ascending order, the first $68\%$ of the models give the range of predicted $w_p$ indicated by the shaded envelope. Our HOD model reproduces the clustering with a reduced $\chi^{2}$ statistic of 0.34 (total $\chi^{2} = 2.39$). Given a $\chi^{2}$ distribution for seven degrees of freedom, the probability of randomly drawing a $\chi^{2}$ value less than or equal to $2.39$ is only $0.065$. This indicates that while the model successfully reproduces the data, the uncertainties quoted on the 2PCF are likely conservative. We discuss this issue further in \S\ref{sec:discussion_fit}. 

Panel (b) shows the mean occupation function from the best-fit HOD model, decomposed into its central (dashed line) and satellite (dot-dashed line) components. Similarly to panel (a), the shaded regions indicate the range of the mean occupation function given by the $68\%$ of models with the smallest $\chi^2$ statistic. The panel shows that while reasonable constraints can already be obtained on the central occupation with current data, only an upper limit can presently be obtained on the satellite occupation. This upper limit suggests that, at low redshift, typically only the most massive halos ($>10^{14} \; \hMsun$) host multiple AGN.  It also yields an upper limit on the satellite fraction of X-ray AGN of $\lesssim0.1$ (at $1\sigma$, $f_{\rm sat}=0.9^{+2.2}_{-0.7} \times 10^{-2}$), where the satellite fraction is defined as the ratio of the number density of satellite AGN (integrated over all halo masses) to the total number density of all AGN. However, we note that this interpretation is sensitive to our choice of satellite parameterization. Alternative models of the mean occupation function \citep[e.g.,][]{k&o12} allow multiple AGN to reside in low-mass halos.

Panel (c) shows the full host halo mass distribution (averaged over all points in the MCMC chain) for central AGN. For each model in the MCMC chain, the distribution is derived by multiplying the mean occupation function of central AGN, $\langle N_{\rm{cen}}(M)\rangle$, with the differential halo mass function. For a randomly chosen central AGN, the curve is the probability density function for the mass of its host halo. Finally, panel (d) shows the probability density function for the median mass of halos hosting central AGN. At the $68\%$ confidence level, denoted by the dotted vertical lines, we find the median host halo mass for central AGN to be $1.02^{+0.21}_{-0.23} \times 10^{13} \; \hMsun$. We do not depict the corresponding distributions for satellite AGN in this figure---the satellite mean occupation is insufficiently constrained for its halo mass distribution to be robustly determined. The values of our best-fit HOD parameters are as follows: $ \log [M_{\mathrm{min}}/(\hMsun)] = 13.65^{+0.1}_{-0.05}$, ${ \sigma_{\log M}} = 0.78^{+0.09}_{-0.06}$, $\log [M_{1}/(\hMsun)] = 14.32^{+7.87}_{-2.15}$, and $\alpha = 2.59^{+0.33}_{-1.87}$. We will discuss these HOD constraints further in \S\ref{sec:discussion_theory}.

The central mean occupation function shown in panel (b) of Figure~\ref{fig:fig1} can also be interpreted as the halo mass-dependent duty cycle of AGN (i.e., the fraction of halos hosting a central AGN). We estimate the duty cycle $f_{\mathrm{o}}$ of central AGN around the median host halo mass. For each model in the MCMC chain, we average the predicted $\langle N_{\rm{cen}}(M)\rangle$ over the mass interval spanned by the central $68\%$ of the median host halo mass distribution (shown in panel d of Figure~\ref{fig:fig1}). From the normalized distribution of these mass-averages we infer, at the $68\%$ confidence level, an average duty cycle of $f_{\mathrm{o}}=0.12 \pm 0.02$ for $z\sim 1.2$ X-ray AGN.

\subsection{Error Forecasting}
\label{sec:results_forecast}

%%%%%%%%%%%%%%%%%%%%%%%%%%%%%%%%%%%%%%%%%%%%%%%%%
\begin{figure*}
\begin{center}
\begin{tabular}{c}
\includegraphics[width=16cm]{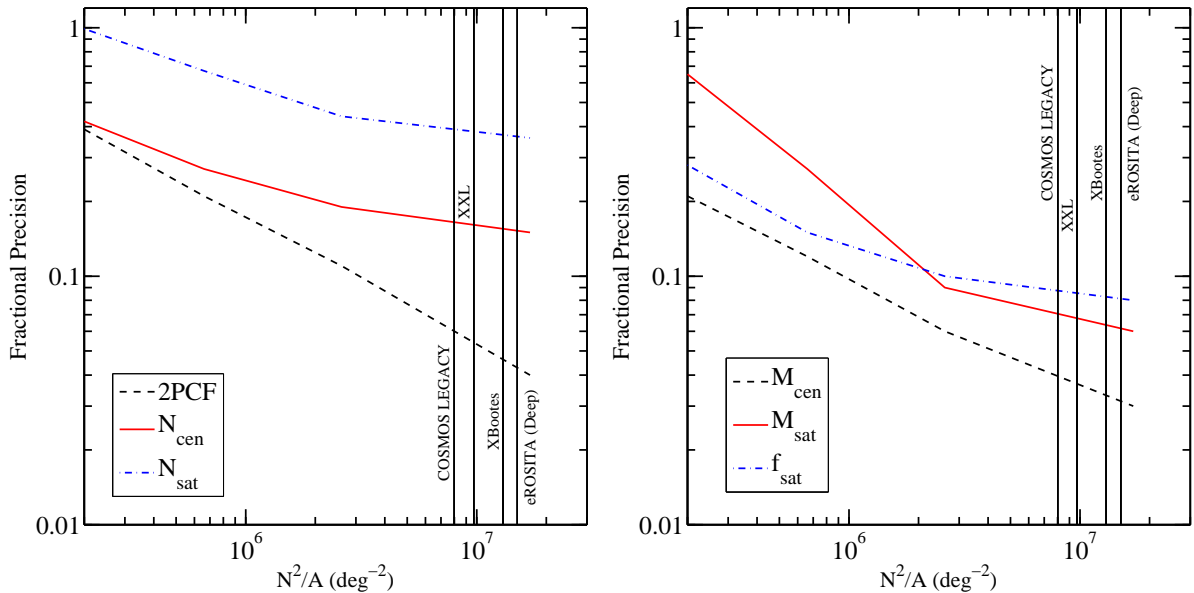}
\end{tabular}
\caption{
\label{fig:fig2}
The forecasted HOD modeling precision obtainable as a function of the survey $N^2/A$ ratio, where $N$ is the number of sources observed in the survey and $A$ is the survey area. The left panel shows the average fractional uncertainties (see the definition in the text) on the 2PCF (dashed line), central mean occupation function (solid line), and satellite mean occupation function (dot-dashed line). The right panel shows the fractional uncertainties on physical parameters derived from the mean occupation function: the median host halo mass for central (dashed line) and satellite (solid line) AGN and the AGN satellite fraction (dot-dashed line). In both panels, the vertical lines denote the $N^2/A$ value of a particular proposed X-ray AGN survey, as labeled  (see Table~\ref{tab:tab1}). The ``XLL'' and ``CMS'' abbreviations refer to the {\it XMM}-XLL and {\it Chandra} 10 deg$^2$ surveys, respectively. Our error forecasting indicates that a large improvement in HOD modeling precision can be realized by planned AGN surveys of modest size. See \S\ref{sec:results_forecast} and \S\ref{sec:survey_design} for additional discussions regarding the feasibility of different planned surveys.
}
\end{center}
\end{figure*}
%%%%%%%%%%%%%%%%%%%%%%%%%%%%%%%%%%%%%%%%%%%%%%%%%%

Although the physical constraints obtained in \S\ref{sec:results_current} are consistent with existing estimates, the current precision of 2PCF measurements does not strongly constrain the full range of possible HOD models.

In order to forecast the improvement in HOD modeling precision that can be achieved with planned X-ray surveys, we construct and model a set of simulated 2PCF measurements based on the original measurements of \citet{allevatoetal11}, but with appropriate rescaling of the error bars. Since the observational AGN number density varies widely between different proposed surveys, an appropriate $\delta w_p$ scaling relation must account for both survey coverage and depth. As will be discussed in \S\ref{sec:survey_design}, for example, a shallow, large-area survey can have a surface density (and thus volume density) an order of magnitude smaller than those of deeper surveys. This will significantly impact the observed number of AGN pairs, which determines the overall precision of the 2PCF measurements ($\delta w_p \appropto 1/\sqrt{N_{\rm pairs}}$, with $N_{\rm pairs}$ the number of pairs). If we assume the redshift distribution to be similar for every proposed survey, the number (volume) density of a given survey scales approximately as the number of sources per unit area. Assuming populations with the same clustering, the number of AGN pairs would then scale as the product of the total number of AGN in the survey, $N$, and the observed AGN surface density, $N/A$, where $A$ is the solid angle of the survey.

We thus rescale our 2PCF error bars according to $1 / \sqrt{N^{2}/A}$ to forecast future HOD constraints. For each $r_p$ bin, we take the corresponding $w_p$ value to be the value of the best-fit 2PCF shown in panel (a) of Figure~\ref{fig:fig1} (i.e., the predicted 2PCF of the HOD model best fitting the \citealt{allevatoetal11} data). We use the best-fit 2PCF values instead of the original measurements to obtain a ``smoothed'' data set free of outliers, since artificially reducing the error bar on an outlying data point can potentially cause the HOD modeling to fail. We find that the pair separation range probed by the \citet{allevatoetal11} sample, $0.5 \; \hMpc < r_p < 30 \; \hMpc$, is insufficient to tightly constrain the satellite occupation, irrespective of the measurement precision. 

Fully constraining the satellite occupation requires measuring the 2PCF to scales well below $\sim 1 \; \hMpc$, the typical halo diameter (e.g., \citealt{richardsonetal12} suggest that 2PCF measurements to $\sim 0.01 \; \hMpc$ scales are needed). Fortunately, larger samples will produce more small-scale pairs, allowing measurement of the 2PCF to smaller scales than was possible with the \citet{allevatoetal11} sample. Hence, in order to isolate the uncertainty in the satellite occupation due to measurement precision, we extend our dummy 2PCF to smaller scales. This measurement will be limited by the angular resolution of the telescope. If we use only the inner region of the {\em Chandra} survey (off-axis angle $\la 7\arcmin$) where the angular resolution is $\la 5\arcsec$ (the full width at half maximum of the point spread function), the smallest resolvable pair separation at $z\sim1$ is $\sim 50 \; \hkpc$ (comoving), which sets the smallest scale in our investigation.

We thus add radial bins centered on 0.1 and 0.05 $\hMpc$ to estimate the satellite constraints. We expect 2PCF measurements to these scales to be achievable by all of the proposed surveys discussed herein, with the exception of the eROSITA mission. Since the sky-averaged eROSITA point spread function is expected to be $\sim30''$, corresponding to $\sim 0.2 \; \hMpc$ at $z\sim0.5$, we note as a caveat that our error forecasting may overestimate the constraints obtainable on the satellite occupation for this particular survey. As with the larger radial bins, we assign to each new bin the $w_p$ value predicted by the HOD model best fitting the \citet{allevatoetal11} data. For a small-scale bin centered on $r_p$ with upper and lower boundaries at $r_{p2}$ and $r_{p1}$, respectively, the uncertainty scales approximately as
\begin{equation}
\delta w_p \propto w_p / \sqrt{\pi(r_{p2}^2-r_{p1}^2)(2\pi_{\rm max} + w_p)},
\end{equation}
where $\pi_{\rm max}$ is the maximum line-of-sight separation between AGN pairs. The constant of proportionality can be determined by matching this relation to the existing \citet{allevatoetal11} measurements at $r_p < 1 \; \hMpc$.

We now define a statistic to describe the overall precision of a given data set in a single number, to which we will refer as the ``average fractional uncertainty.'' Applied to the projected 2PCF measurements, the average fractional uncertainty is calculated by computing the fractional error on each data point, $\delta w_{p}(r_p)/w_p(r_p)$, and then averaging the fractional errors over all $r_p$ bins. This statistic is calculated similarly for the central and satellite mean occupation functions, $\langle N_{\mathrm{cen}}(M)\rangle$ and $ \langle N_{\mathrm{sat}}(M)\rangle$, respectively, but the halo mass range is restricted to the interval over which the mean occupation function is greater than $10^{-2}$. We impose this condition so as to consider the uncertainty only over the physical halo mass range (i.e., the range that contributes significantly to the AGN number density).

Figure~\ref{fig:fig2} shows the forecasted HOD modeling precision obtainable as a function of the survey $N^2/A$ ratio. The left panel shows the average fractional uncertainties on the 2PCF (dashed line), central mean occupation function (solid line), and satellite mean occupation function (dot-dashed line). The right panel shows the fractional uncertainties on physical parameters derived from the mean occupation function: the median host halo mass for central (dashed line) and satellite (solid line) AGN and the AGN satellite fraction (dot-dashed line). In both panels, the vertical lines denote the $N^2/A$ value of a particular proposed X-ray AGN survey, as labeled. Our error forecasting indicates that a large improvement in HOD modeling precision can be realized by planned AGN surveys of modest size. For a survey of area equal to that of the {\it XMM}-COSMOS field ($2.13$ deg$^{2}$), for example, we find that a sample of $\sim 1,000$ AGN with spectroscopic redshifts will be sufficient to double the precision of current constraints, and a sample size larger than 10,000 will allow determination of the derived physical parameters to to better than $10\%$ statistical uncertainty. In the next section we discuss the plausibility of current and future X-ray surveys in achieving this precision.

\subsection{Survey Design}
\label{sec:survey_design}

%%%%%%%%%%%%%%%%%%%%%%%%%%%%%%%%%%%%%%%%%%%%%%%%%%%%%%%%%%%%%%%%%%%%%%%%%%%%%%%%%%%%%%%%%%%%%%%%%%%%%%%%%%%%%%%%%%%%%%%%%%%%%%%%%%%%%%%%%%%%%%%%%%%%%%5
\begin{deluxetable*}{lcccccccrc}
\tablecaption{
\label{tab:tab1}
AGN source counts in X-ray surveys
}
\tablehead{
\colhead{} &
\colhead{Area ($A$)} &
\colhead{$f_{20}$} &
\colhead{$f_{80}$} &
\multicolumn{3}{c}{Number($N$)$/ \log[L_{\rm X}/ (\ergs)]$ bin} &
\colhead{$N/A$} &
\colhead{$N^{2}/A$} &
\colhead{} \\
\colhead{Survey} &
\colhead{(deg$^2$)} &
\colhead{(erg cm$^{-2}$ s$^{-1}$)} &
\colhead{(erg cm$^{-2}$ s$^{-1}$)} &
\colhead{42--43} &
\colhead{43--44} &
\colhead{44--45} &
\colhead{(deg$^{-1}$)} &
\colhead{(deg$^{-2}$)} &
\colhead{Ref}}
\startdata
{\em Chandra} COSMOS Legacy & 2 & $2.0\times10^{-16}$ &
$3.7\times10^{-16}$ & 1600 & 2100 & $460$ & 2000 & $8\times10^{6}$ & (1) \\
{\em Chandra} 10 deg$^2$ survey & 9 & $6.1\times10^{-16}$ &
$7.9\times10^{-16}$ & 2900 & 6300 & $1900$ & 1200 & $1.3\times10^{7}$ & (2) \\
{\em XMM}--XXL & 50 & $3.1\times10^{-15}$ & $1.0\times10^{-14}$ & 1300
& 5800 & 4000 & 440 & $9.7\times10^{6}$ & (3) \\
eROSITA (Deep) & 100 & $3.1\times10^{-15}$ & $3.2\times10^{-15}$ &
4700 & 21000 & $12000$ & 380 & $1.5\times10^{7}$ & (4) \\
eROSITA (All Sky) & 41253 & $10^{16}$ & $10^{-16}$ & 320000 &
1600000 & $2000000$ & 95 & $3.7\times10^{8}$ & (5)\\
\enddata
\tablecomments{
The values in columns $N/A$ and $N^2/A$ are calculated over all luminosity bins ($42 < \log[L_{\rm X}/ (\ergs)] < 45$). References: 1. F. Civano (private comm.), 2. Scaled from \citet{kenteretal05}, 3. \citet{elyivetal12}, \citet{pierre12} 4. \citet{merlonietal12}, 5. \citet{merlonietal12}.
}
\end{deluxetable*}
%\end{center}

%\footnotetext{$}
%%%%%%%%%%%%%%%%%%%%%%%%%%%%%%%%%%%%%%%%%%%%%%%%%%%%%%%%%%%%%%%%%%%%%%%%%%%%%%%%%%%%%%%%%%%%%%%%%%%%%%%%%%%%%%%%%%%%%%%%%%%%%

Based on our error forecasts we now evaluate our predictions in light of ongoing and future surveys of X-ray AGN. In Table~\ref{tab:tab1} we show the relevant source statistics. The number counts are obtained using the predictions of the cosmic X-ray background model of \citet{gillietal07}\footnote{http://www.bo.astro.it/~gilli/counts.html} and sensitivity curves for various X-ray surveys. All X-ray luminosities and fluxes are quoted in the 0.5-2 keV band. The counts are for sources detected at all redshifts (the redshift distribution peaks at $z\sim 1$ independent of flux limit). The parameters $f_{20}$ and $f_{80}$ represent the flux limits at 20\% and 80\% of the total survey area, respectively. These two numbers are useful in providing a more accurate shape of the sensitivity curve. In column 8 of Table~\ref{tab:tab1} we compute the approximate source densities (number of yield per unit area) of relevant X-ray surveys. It is important to note that the larger area, shallower surveys have smaller surface densities (and thus volume densities) than the deeper surveys. This potentially affects the number of pair counts in clustering studies.

We find that the {\em Chandra} COSMOS Legacy Survey, the {\em XMM}--XXL Survey, and a {\em Chandra} medium-area ($\sim 10$ deg$^2$) survey can all obtain high-precision HOD constraints. However, as mentioned previously, our requirement that spectroscopic redshifts be obtained for our sources may prove challenging for a $50 \; {\rm deg}^{2}$ survey like {\em XMM}--XXL. Hence we emphasize that a deeper survey covering a smaller area is preferable for HOD studies, given that the surface density of sources is larger in this case and follow-up spectroscopy is better facilitated\footnotemark. It is evident from Table~\ref{tab:tab1} that either the COSMOS Legacy Survey or a medium-area {\em Chandra} survey can be utilized to study luminosity dependence of the HOD. The \citet{allevatoetal11} clustering sample has been constructed over a wide luminosity range ($10^{41} -10^{45} \; \ergs$) and our HOD precision estimates in Figure~\ref{fig:fig2} are obtained using the median-redshift/median-luminosity interpretation. While these precision forecasts are robust in terms of statistical uncertainties, with increasing statistical precision in the 2PCF, systematic effects from luminosity and/or redshift dependence will become increasingly important and may limit the ultimate precision of constraints obtainable on the HOD (i.e., the median-redshift/median-luminosity interpretation may no longer apply; see \S\ref{sec:discussion_fit}). In this case, more sophisticated HOD models with parameters accounting for redshift and/or luminosity dependence would be warranted.

In both panels of Figure~\ref{fig:fig2}, all of the fractional precision curves for the HOD modeling scale (roughly) linearly with the precision curve for the 2PCF before flattening once it surpasses $\sim 10\%$ precision (near $N^2/A \sim 2 \times 10^6 \; {\rm deg^2}$). The characteristic shape of these curves arises from a transition in the dominant source of statistical uncertainty in the modeling. As the fractional uncertainty of the 2PCF decreases with increasing $N^{2}/A$, it eventually falls below the fractional uncertainty of the AGN number density, which we have fixed to $10\%$. We investigated a wide range of number densities and fractional uncertainties (up to a factor of $\sim 5$ variation) to determine their impact on our HOD modeling results. In all cases, we recovered similar curves to those shown in Figure~\ref{fig:fig2} which flattened near the $N^{2}/A$ value of 2PCF-number density fractional uncertainty equality. Since the planned surveys discussed herein all fall within the number density-limited regime ($N^2/A > 2 \times 10^6 \; {\rm deg^2}$), we note that their forecasted precisions are limited by the assumed $10\%$ uncertainty in the number density, rather than the precision of the 2PCF, and therefore may be conservative estimates. If the systematics of the surveys can be controlled to better than $10\%$ (which we do not assume), it may be possible to obtain slightly tighter constraints on the HOD than those indicated here.

Finally, we note that the survey error forecasts presented herein are calculated for AGN {\it auto}-correlation measurements. Although it is beyond the scope of the current work, some studies have suggested that even higher precision clustering constraints may be achievable through the cross-correlation between AGN and galaxies (see \citealt{hickoxetal09}, \citealt{coiletal09}, \citealt{krumpeetal10}, \citealt{krumpeetal12} for cross-correlation studies using X-ray AGN) and could possibly lead to HOD constraints if the galaxy population is well-characterized \citep[see, e.g.,][]{wakeetal08, shenetal12a, miyajietal11}.

\footnotetext{Novel statistical techniques have been developed to maximize the precision of clustering measurements using photometric redshift samples \citep{myersetal09, hickoxetal11}, but the statistical power of these measurements is necessarily limited compared to spectroscopic studies using comparable sample sizes.}

\section{Discussion}

We now discuss the systematic issues and theoretical aspects of our analysis. 

\subsection{Systematic Uncertainties}
\label{sec:discussion_fit}

As mentioned previously, we apply the median-redshift/median-luminosity interpretation of \citet{richardsonetal12} to our HOD modeling. Within this interpretation, the HOD can be taken to represent AGN at the median sample redshift if the luminosity and redshift dependence of the 2PCF are weak (i.e., their systematic effects do not exceed the statistical uncertainty quoted on the modeling results). \citet{richardsonetal12} provide a detailed discussion of how intrinsic luminosity or redshift dependence of the clustering would affect the HOD modeling.

To high precision, the clustering of optically-selected quasars has not been found to exhibit significant luminosity dependence \citep[e.g.,][]{croometal05, myersetal07a, shenetal09, shenetal12a}. The luminosity dependence of X-ray AGN clustering at $z \sim 0.25$ has been found to be weak \citep{krumpeetal10, miyajietal11}, and available studies at $z \sim 1$ do not show any evidence for strong dependence \citep[e.g.,][]{coiletal09}. Future studies with larger samples of X-ray-selected AGN are needed to confirm the degree of luminosity dependence at higher redshift. Since the luminosity dependence at $z \sim 1$ has not been well quantified, we assume that the luminosity dependence is weak and is incorporated within the uncertainties of the 2PCF.

There has not been a comprehensive study of the redshift evolution of the clustering of X-ray-selected AGN (due to small sample sizes). However, we examine the redshift evolution of bias from the \citet{allevatoetal11} sample to infer the approximate redshift evolution of the 2PCF. \citet{allevatoetal11} measure the bias of the full sample (median $\langle z \rangle =1.2$) to be $b = 2.98 \pm 0.13$. They then construct a sub-sample spanning a smaller redshift interval, but with a similar median redshift ($\langle z \rangle =1.3$) to that of the full sample, and evaluate the bias of the sub-sample to be $b = 3.10 \pm 0.18$. The consistency between the clustering amplitudes over the whole redshift range and in the narrow redshift range around the median redshift therefore lends supports to our interpretation of the HOD modeling results.

\citet{richardsonetal12} provide a thorough discussion on the implications of alternative HOD models on the quasar 2PCF. In short, \citet{richardsonetal12} found the central occupation to be robust to the chosen HOD parameterization but found the satellite occupation to exhibit statistically significant model-dependences at high halo masses. This systematic effect led to a statistically significant difference in some physical parameters. For example, the quasar satellite fractions obtained from HOD modeling by \citet{k&o12} and \citet{richardsonetal12} disagreed with each other by a large factor (two substantially different HOD models were assumed by the two different groups). Although breaking this degeneracy in the high-mass satellite occupation will require larger samples of small-separation pairs coupled with additional constraints (e.g., the distribution of line-of-sight pair velocities), it is not of central concern to the current X-ray AGN constraints presented herein. Our modeling is unable to impose strong constraints on the form of the satellite occupation due to the lack of small-scale clustering data ($< 1 \; \hMpc$).

As an additional check, we generalize the form of our central HOD parameterization by allowing the softened step function to approach an arbitrary asymptotic value $f$ less than unity (see Eq.\ 1 and the parameterization of \citealt{allevatoetal12}) and repeat the modeling of the \citet{allevatoetal11} data. We find that the HOD fit is statistically insensitive to this new degree of freedom, and we recover similar relative uncertainties to those yielded by the original model. The central and satellite occupation functions remain consistent at the $1 \sigma$ level with our original modeling. The more flexible model also does not lead to a statistically significant change (at the $1\sigma$ level) in any of the physical parameters derived from the HOD, including the upper limit on the satellite fraction (at $1\sigma$, $f_{\rm sat}=0.8^{+2.9}_{-0.6} \times 10^{-2}$). The best-fit HOD parameters obtained for this alternate model are as follows: $ \log [M_{\mathrm{min}}/(\hMsun)] = 13.01^{+0.44}_{-0.25}$, $ \sigma_{\log M} = 0.45 \pm 0.28$, $f = 0.83^{+0.17}_{-0.75}$, $\log [M_{1}/(\hMsun)] = 14.98^{+1.39}_{-0.57}$, and $\alpha =  2.16^{+0.88}_{-0.85}$.

Although our chosen model has been calibrated for low-luminosity AGN, we emphasize that the current model has successfully reproduced quasar clustering and has been favored by direct measurements of the mean occupation function of X-ray AGN \citep{allevatoetal12}. We thus use the model proposed by \citet{chatterjeeetal12} to directly compare the HOD of X-ray AGN to that of luminous quasars. We also compare our HOD constraints obtained from modeling the 2PCF to those directly inferred for X-ray AGN by \citet{allevatoetal12} in \S\ref{sec:discussion_theory}.

Finally, while our HOD model successfully reproduces the observed clustering of X-ray AGN, it fits the data with a reduced $\chi^2$ statistic of 0.34. This suggests either that our HOD model admits too many free parameters, thereby overly fitting the details in the 2PCF, or that the error bars quoted on the 2PCF overestimate the true $1\sigma$ uncertainty. To test whether the four-parameter model overfits the data, we fit only the central occupation function to the two-halo term (all data points at $r_p > 1 \; \hMpc$) with the parameter $\sigma_{\rm logM}$ fixed to the best-fit value yielded by the full modeling. Hence we only fit a single free parameter, $M_{\rm min}$, to nine data points. Even with most of the freedom of the HOD model eliminated, the modeling still reproduces the data with a reduced $\chi^2$ statistic of $\sim 0.4$. We interpret this to indicate that the bootstrap errors estimated by \citet{allevatoetal11} are likely a conservative estimate of the true uncertainty. As a caveat to our modeling, however, we note that neglecting the covariance in the 2PCF measurements can also potentially lead to an overfitting of the data.

\subsection{Comparison with Previous Work}
\label{sec:discussion_theory}

%%%%%%%%%%%%%%%%%%%%%%%%%%%%%%%%%%%%%%%%%%%%%%%%%%
\begin{figure}[t]
\begin{center}
\begin{tabular}{c}
\includegraphics[width=8cm]{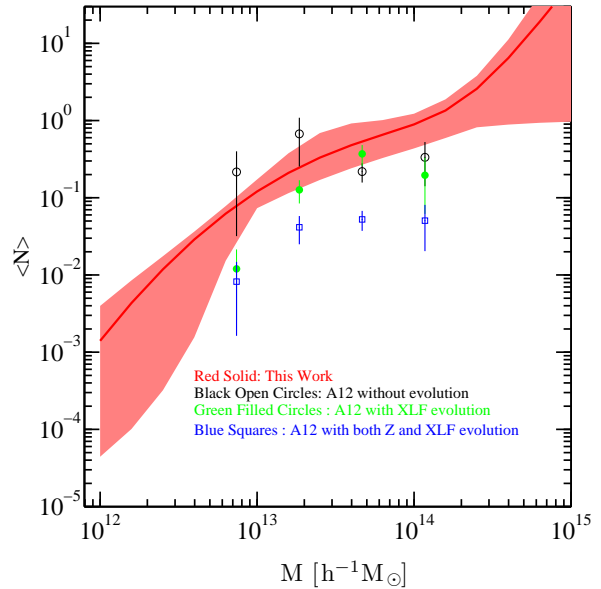}
 \end{tabular}
 \caption{
 \label{fig:fig3}
Comparison of the total mean occupation function of X-ray AGN as empirically measured with a group catalog by \citeauthor{allevatoetal12} (\citeyear{allevatoetal12}; black open circles) to the 2PCF modeling inference from the current work (red solid line). The red shaded region indicates the $68\%$ confidence interval of our modeling (see the text). The green filled circles and the blue open squares represent the empirically-measured mean occupation function of \citet{allevatoetal12} after correcting for luminosity dependence and redshift $+$ luminosity dependence of the X-ray AGN, respectively. See \S\ref{sec:discussion_theory} for further discussion of the comparison between the two measurements.
}
\end{center}
\end{figure}
%%%%%%%%%%%%%%%%%%%%%%%%%%%%%%%%%%%%%%%%%%%%%%%%%%

Our analysis suggests that the $68\%$ confidence interval for the best-fit host halo mass of central X-ray AGN is $12.90 \leq \log [{\rm M_{cen}}/(h^{-1} \; {\rm M}_{\odot})] \leq 13.09$. Using a simple HOD model consisting of only central AGN and characterized by a delta function for the occupation distribution, \citet{allevatoetal11} obtained a host halo mass of ${\rm \log [M_{cen}}/(h^{-1} \; {\rm M}_{\odot})] = 12.97 \pm 0.06$, which is consistent with our result. Previously, \citet{coiletal09} measured the CCF of X-ray AGN selected from the All-Wavelength Extended Groth Strip International Survey with a control sample of galaxies and obtained a minimum host halo mass of $5^{+5}_{-3} \times 10^{12} \; h^{-1} \; {\rm M}_{\odot}$, which is also consistent with our result. The median redshift and luminosity of their AGN sample, $0.90$ and $10^{42.8}\;\ergs$, respectively, are similar those of the \citet{allevatoetal11} sample.

A related study was carried out by \citet{starikovaetal11}. Using a sample of X-ray-selected AGN in the {\em Chandra} Bo\"{o}tes field ($41 \leq \log [L_{\rm X}/(\ergs)] \leq 45$), they performed HOD modeling by comparing the correlation function projected parallel and perpendicular to the line of sight. For $0.17 \leq z \leq 3.0$, their analysis preferred a host halo mass scale $ > 4.1 \times 10^{12} \; \hMsun$ and found that AGN reside primarily in central galaxies. The derived host halo mass scale is consistent with our results but our satellite occupation is insufficiently constrained for a quantitative comparison of satellite fractions.

Recently, \citet{allevatoetal12} used a novel approach to directly measure the occupation function of X-ray AGN in groups and clusters. In Figure~\ref{fig:fig3} we compare our total mean occupation function obtained from modeling the 2PCF (solid red line) to the direct measurement of \citet{allevatoetal12}\footnotemark (black open circles). The error-bars correspond to the $1\sigma$ errors in mass and occupancy and the red shaded region represents the $68\%$ confidence interval of our modeling. It is evident from Figure~\ref{fig:fig3} that the two complimentary measurements are consistent with each other, thus validating the HOD parameterization that we have used in the 2PCF modeling. As previously mentioned, the model was developed using numerical simulations of lower luminosity AGN and has been applied to the current sample under the assumption of universality in the general AGN HOD properties.

However, the populations of X-ray AGN in the two samples are not exactly the same. The \citet{allevatoetal12} sample consists of lower-redshift AGN ($z < 1.0$) with slightly lower X-ray luminosities compared to the \citet{allevatoetal11} sample, whose 2PCF measurements we model in the current work. \citet{allevatoetal12} attempt to correct their measurement of the mean occupation function for luminosity and redshift dependence. The data points, represented by the green filled circles and the blue open squares in Figure~\ref{fig:fig3} signify the mean occupation functions after correcting for luminosity dependence and luminosity $+$ redshift dependence, respectively. We find a disagreement ($1.7 \sigma$ level) between the two measurements once the direct measurement is corrected for redshift evolution. Physically, our $\langle N(M) \rangle$ can be interpreted as the mass-dependent AGN occupation fraction at $z \sim 1.2$, under the assumption of weak luminosity and redshift dependence (see \S\ref{sec:discussion_fit}).  While it is possible that this disagreement in the occupation fraction could arise due to luminosity and/or redshift dependence of the X-ray AGN population, we emphasize that it could also arise merely from an over- or under-correction for dependence. Conclusively identifying the source of this discrepancy will require studies of redshift and luminosity dependence with larger AGN samples.

\footnotetext{There is a typographical error in the coefficients of the ${L_{\rm X}-M}$ scaling relation in \citet{allevatoetal12}. The correct coefficients \citep{leauthaudetal10} have been adopted in determining the masses of the X-ray groups and clusters in the paper (V. Allevato, private comm.).}

\subsection{Implications for AGN Evolution}
\label{sec:discussion_implications}

%%%%%%%%%%%%%%%%%%%%%%%%%%%%%%%%%%%%%%%%%%%%%%%%%%
\begin{figure*}[t]
\begin{center}
\begin{tabular}{c}
 \includegraphics[width=16cm]{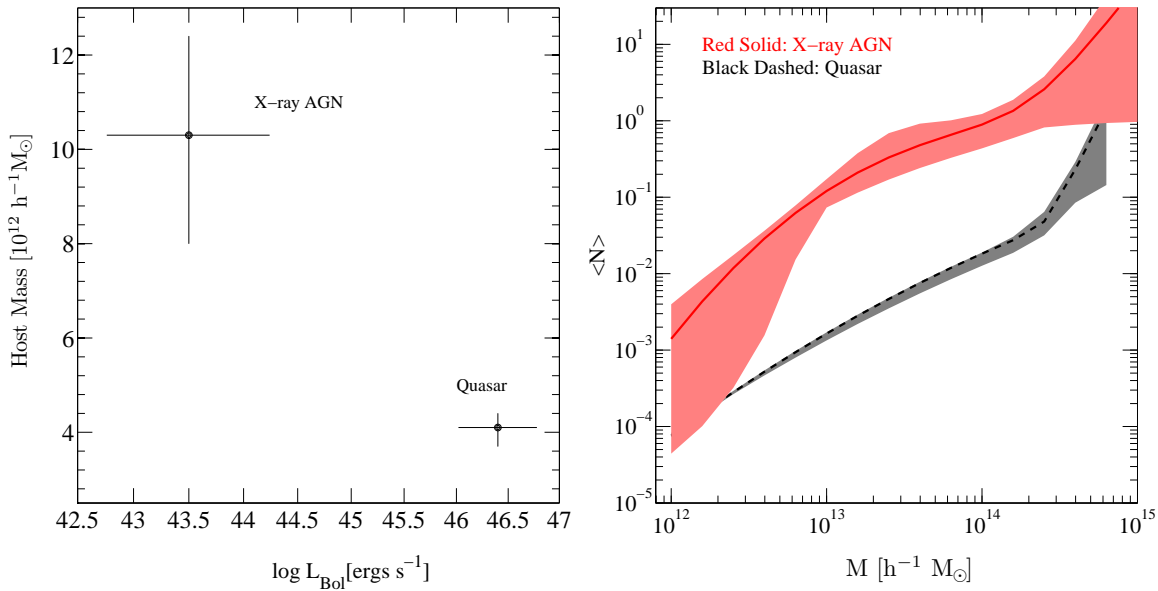}
\end{tabular}
\caption{
\label{fig:fig4}
Left panel: the median halo mass of quasars and X-ray AGN as a function of bolometric luminosity at $z \sim 1$, as labeled. The quasar result is obtained from \citet{richardsonetal12} and the X-ray AGN constraint is from the current work. Both values are obtained from HOD modeling of the 2PCF. This shows that there is a $2.5\sigma$ difference between the host halo masses of quasars and X-ray AGN, with the latter residing in higher mass halos. This result is similar to the evolutionary picture presented in \citet{hickoxetal09} for AGN at $z \sim 0.5$, but now performed at significantly higher redshift (see also \S\ref{sec:discussion_implications}). Right panel: the predicted total mean occupation function for quasars (dashed line; \citeauthor{richardsonetal12} \citeyear{richardsonetal12}) compared to the predicted total mean occupation function of X-ray AGN (solid line; current work).
}
\end{center}
\end{figure*}
%%%%%%%%%%%%%%%%%%%%%%%%%%%%%%%%%%%%%%%%%%%%%%%%%%

From the measured clustering of AGN at different wavelengths and at different redshifts one can draw an evolutionary picture of AGN growth and connect it to the cosmic growth of structure (popularly known as `AGN co-evolution' in the literature). We now interpret our HOD result within this co-evolution paradigm. We adopt the evolutionary picture presented in \citet{hickoxetal09}, which has been successful in interpreting several theoretical aspects of the co-evolution scenario, to assess our findings. We will focus on the results of \citet{richardsonetal12} for the quasar HOD at $z \sim 1.4$ and the results of the current paper for the X-ray AGN HOD at $z \sim 1.2$. Critically, the X-ray AGN we consider are substantially less bolometrically luminous than quasars. The \citet{hickoxetal09} picture has been evaluated primarily on the basis of moderate-redshift observations ($z \sim 0.5$). We now extend this picture to a higher redshift ($z \sim 1$).

In this picture, massive galaxies follow an evolutionary sequence from a gas-rich star-forming quasar phase, to a spheroid with lower star formation and AGN activity, to a red-and-dead galaxy with discrete radio outbursts from the central AGN \citep[e.g.,][]{crotonetal06, hopkinsetal08, hickoxetal09}. The gas-rich quasar phase could be possibly driven by major mergers \citep[e.g.,][]{k&h00, hopkinsetal06a, hopkinsetal08} or by secular instabilities \citep[e.g.,][]{boweretal06}. It is believed to occur at a typical halo mass scale of a few times $10^{12} \; {\rm M}_{\odot}$ at almost all redshifts and may result from a natural evolution of the halo-galaxy mass relation, given that galaxy formation is most efficient at those halo mass scales \citep[e.g.,][]{c&w13} and powerful starbursts galaxies are found in halos of similar mass (see \citeauthor{hickoxetal12} \citeyear{hickoxetal12} and references therein). In this phase the central black hole undergoes Eddington accretion and shines as a luminous quasar.

From the quasar phase galaxies transition into the spheroid phase characterized by lower star formation and declining AGN activity. These spheroids reside in halos of mass $\sim 10^{13} \; {\rm M}_{\odot}$. Their central black holes undergo moderate accretion ($0.001-0.1$ Eddington) and shine as X-ray-bright AGN. The cause of this decline in AGN activity is not yet known, but a popular model involves a lack of cold gas due to shock heating inside the halo \citep[e.g.,][]{b&d03, keresetal09} or feedback from the AGN itself \citep[e.g.,][]{s&r98,dimatteoetal05}. At even later times, galaxies enter the red-and-dead phase in which the central AGN becomes extremely radiatively inefficient. Continued growth of the host halo completely shock-heats the intra-halo gas. From clustering studies \citet{hickoxetal09} demonstrated that X-ray AGN reside in dark matter halos in the mass range of $\sim 10^{13} \; \hMsun$. Using a sample of radio AGN, which have very low accretion efficiencies, \citet{hickoxetal09} also demonstrated that radio AGN are found in halos of mass $\sim 3\times 10^{13} \; \hMsun$ or greater, consistent with the results of several other studies \citep[e.g.,][]{wakeetal08,mandelbaumetal09,donosoetal10}

\citet{richardsonetal12} found the median host halo mass of central quasars at $z \sim 1.4$ to be $4.1^{+0.3}_{-0.4} \times 10^{12} \; \hMsun$, which is significantly ($2.5 \sigma$) lower than the corresponding host halo mass scale of X-ray-selected AGN near this redshift. The left panel of Figure~\ref{fig:fig4} shows the respective locations of quasars and X-ray AGN in (bolometric) luminosity-host halo mass space. The errors represent the FWHM around the median bolometric luminosities of the sample quasars and X-ray AGN. It is apparent that, at $z \sim 1$, X-ray AGN with low bolometric luminosities reside in halos of significantly higher mass than quasars with bolometric luminosities at least two orders of magnitude higher. Since, even at higher redshifts, higher luminosity AGN (i.e., quasars) reside in halos of lower mass compared to lower luminosity AGN, we confirm that the correlation between AGN luminosity and host dark matter halo mass is relatively weak. Moreover, black holes indeed appear to undergo the quasar phase near the critical mass limit of $\sim 10^{12} \; {\rm M}_{\odot}$ and generally become less active as the halo grows in mass. Thus our detailed HOD modeling supports the \citet{hickoxetal09} picture at high redshift. 

We note that the ``cartoon'' picture presented in \citet{hickoxetal09} is valid only for {\em central galaxies}. The fact that our HOD modeling shows that AGN reside primarily in central galaxies validates interpreting the AGN population within this picture. An HOD analysis of the 2PCF of high-redshift radio-selected AGN could further constrain this evolutionary picture. Some studies have suggested that, at high redshift, the CCF of radio-loud quasars with their radio-quiet counterparts does show more clustering strength than the auto-correlation function of radio quiet quasars \citep[e.g.,][]{shenetal09}, which would indicate that radio-loud quasars reside in more massive halos. It is important to note that the HOD and black hole properties of AGN are likely to depend on several other properties of the AGN. For example it has been shown by \citet{krumpeetal12} that at low redshifts ($z < 1$) the clustering of X-ray-selected and optically-selected broad line AGN are statistically identical. It has been further shown in this study that only narrow line radio-loud AGN exhibit enhancement in clustering strength compared to their radio-quiet counterparts.

From the average duty cycle measurements of quasars and X-ray AGN we can infer a characteristic lifetime of these objects. The lifetime can be approximately estimated as $t_{o} = f_{o}\times t_{\rm H}$ \citep{m&w01,croton09,h&h09}, where $t_{\rm H}$ is the Hubble time and $f_{o}$ is the average duty cycle. For our purpose, the duty cycle can be estimated as the occupation fraction of AGN or quasars at the median redshift, which is shown in the right panel of Figure~\ref{fig:fig4}. For quasars, \citep{richardsonetal12} obtained an average duty cycle of $f_{\mathrm{o}}=7.3^{+0.6}_{-1.5}\times 10^{-4}$ at $z\sim 1.4$. The Hubble time is roughly $6$ Gyr at $z \sim 1$, yielding a characteristic quasar lifetime of $\sim 4.4$ Myr. Similarly, from the HOD modeling results of the current work, we find a characteristic X-ray AGN lifetime of $\sim 0.7$ Gyr. Since these lifetimes invoke duty cycles calculated for the median host mass scale, we interpret these timescales as the mean lifetime of a ``typical'' quasar or AGN at $z\sim1$.

We can compare our estimates of lifetime with the semi-analytic predictions of \citet{hopkinsetal09}. Using an observed distribution of Eddington ratios and AGN model light curves, \citet{hopkinsetal09} found that a black hole spends different amounts of time at different stages of its evolutionary sequence. The lifetime at high Eddington ratios ($\geq 0.1$) is typically $10-100$ Myrs and at moderate Eddington ratios ($0.001 - 0.1$) is typically $0.5-1$ Gyr. Our current sample of X-ray-selected AGN are likely to have moderate Eddington ratios \citep[e.g.,][]{hickoxetal09} while the quasar population is representative of the high accretion stages of a black hole. Our estimated lifetimes, both for the highly accreting quasar phase and the moderately accreting X-ray phase, are in broad agreement with the \citet{hopkinsetal09} prediction. As we do not have a catalog of the masses of our black holes, this comparison is necessarily mostly qualitative in nature. It has been discussed in \citet{hopkinsetal09} that the lifetimes of black holes are a function of several parameters (e.g., black hole mass, luminosity, host halo mass). Hence a more accurate comparison of lifetimes with the \citet{hopkinsetal09} picture is possible with black hole mass measurements of our quasar and X-ray AGN samples. Qualitatively, our results independently support the idea that black holes spend significantly different times in different phases of their evolutionary sequence. On average, we find the lifetime of black holes in the moderate Eddington phase to be $\sim100$ times longer than that of the quasar phase. We summarize the key findings of our analysis in the following section.

\section{Summary of Results}
\label{sec:conclusion}

We perform an HOD analysis of the projected 2PCF of X-ray-selected AGN at $z\sim1.2$. We use a physically-motivated HOD model based on low-luminosity AGN in cosmological simulations, but which has also been found to provide an excellent description of the clustering of quasars \citep{richardsonetal12} and the mean occupation empirically measured for X-ray AGN \citep{allevatoetal12}. Our modeling yields a median mass for halos hosting central AGN of $1.02^{+0.21}_{-0.23} \times 10^{13} \; \hMsun$ and an upper limit on the AGN satellite fraction of $\sim 0.10$, which are consistent with previous estimates. Our 2PCF data sample an insufficient range of the one-halo term to conclusively rule out a monotonically decreasing satellite occupation at high halo mass, but we find that our analysis strongly favors a positive slope.

Since \citet{richardsonetal12} inferred a host halo mass of $4.1^{+0.3}_{-0.4} \times 10^{12} \; \hMsun$ for central quasars at similar redshift, we show that X-ray AGN reside in halos of significantly ($2.5 \sigma$) higher mass compared to quasars at $z \sim 1$.  We also show that at $z \sim 1$, the average lifetime of the X-ray AGN phase is $\sim100$ times longer than that of the quasar phase. The derived mass scales and lifetime estimates support the \citet{hickoxetal09} picture of AGN evolution (established at low redshift), where black holes are believed to follow an evolutionary sequence from a high-Eddington quasar phase to a moderately accreting X-ray phase, to a radiatively inefficient radio phase.

Based on our current analysis, we project the 2PCF measurements of \citeauthor{allevatoetal11} (\citeyear{allevatoetal11}; with appropriate rescaling of the error bars) to forecast the improvement in HOD modeling precision that can be achieved with future X-ray surveys. For a survey of equal area to {\it XMM}-COSMOS, we find that a sample of $\sim 3,000$ AGN is sufficient to constrain the HOD at the $20\%$ level, while a sample size greater than 10,000 will enable characterization of the HOD to at least the $10\%$ level (and possibly better, depending on survey systematics). However, realizing this precision for the satellite occupation and its derived parameters will require measuring the 2PCF to smaller pair separation scales. Based on the results of \citet{richardsonetal12}, we argue that this minimum scale is $\sim 0.01 \; \hMpc$. This should be possible with future X-ray surveys, whose larger volumes will a contain greater number of small-scale pairs. Our projections show that ongoing and proposed X-ray surveys such as the {\em Chandra} COSMOS Legacy survey, a {\em Chandra} medium-area survey, and the eROSITA mission will easily obtain $\sim10\%$-level precision on physical HOD parameters, provided that there is sufficient optical spectroscopic follow-up to obtain accurate redshifts. Accurately determining the AGN HOD will enable key physical parameters like the AGN satellite fraction and duty cycle to be inferred with unprecedented precision, making possible a definitive evaluation of the AGN co-evolution paradigm.

\section*{Acknowledgments}

We thank Viola Allevato and Nico Cappelluti for kindly providing their 2PCF measurements and guidance in using their data set, Yue Shen for useful discussions on the errors on quasar bolometric luminosities and Alexie Leauthaud for her inputs on scatter in the mass measurements of clusters.  We also thank the referee for providing us with several important suggestions which helped in improving the draft. We especially thank Francesca Civano for her help in obtaining the estimates for Table~\ref{tab:tab1}. SC thanks Daisuke Nagai for several useful discussions and acknowledges support from YCAA during the initial phase of the project. SC and ADM were partially supported by the National Science Foundation through grant number 1211112 and by NASA through ADAP award NNX12AE38G and Chandra award number AR0-11018C issued by the Chandra X-ray Observatory Center, which is operated by the Smithsonian Astrophysical Observatory for and on behalf of the National Aeronautics Space Administration under contract NAS8-03060. RCH was supported in part by the National Science Foundation through grant number 1211096 and by NASA through ADAP award NNX12AE38G. JR acknowledges the support and computational resources provided by the University of Chicago Department of Astronomy and Astrophysics through a research assistantship. ZZ is supported in part by NSF grant AST-1208891.

\bibliography{mybib}{}
\bibliographystyle{apj}
\end{document}